\documentclass[%
 aip,
 jmp,%
 amsmath,amssymb,
preprint,%
 reprint,%
author-year,%
author-numerical,%
]{revtex4-2}

\usepackage{graphicx}
\usepackage{dcolumn}
\usepackage{bm}

\begin{document}



\newcommand{\chp}[1]{\mbox{$\stackrel{\wedge}{#1}$}}
\newcommand{\ichp}[1]{\mbox{$\stackrel{\vee}{#1}$}}
\newcommand{\jchp}[1]{\mbox{$\stackrel{\_}{#1}$}}
\newcommand{\ch}[1]{\mbox{$\stackrel{\sim}{#1}$}}
\newcommand{\vct}[1]{\mbox{$\stackrel{\rightarrow}{#1}$}}
\newcommand{\ft}[1]{\mbox{$\stackrel{\wedge}{#1}$}}
\newcommand{\slas}[1]{\mbox{${{#1} \!\!\! /}$}}
\newcommand{\Mob}[0]{M\"{o}bius }



\preprint{AIP/123-QED}

\title[Final state radiation]{Calculation without IR divergence of the soft photon high energy limit of final state radiation for the process $e^{+}e^{-}\rightarrow\mu^{+}\mu^{-}\gamma$}

\author{John Mashford}
 \altaffiliation[ ]{School of Mathematics and Statistics \\
University of Melbourne, Victoria 3010, Australia \\
E-mail: mashford@unimelb.edu.au}

 \homepage{https://findanexpert.unimelb.edu.au/profile/11242-john-mashford}

\date{\today}

\begin{abstract}

In this paper it is shown that the final state radiation process $e^{+}e^{-}\rightarrow\mu^{+}\mu^{-}\gamma$ at tree level is not associated with any IR divergence in the soft photon high energy limit if the calculation is done using a careful treatment of a certain distributional object (in fact, a measure) arising from the Feynman amplitude for the process. Thus there need be no ``infrared catastrophe" associated with the process. It is shown that, in fact, the cross section for the final state photons for the the process vanishes in the soft photon high energy limit.

\end{abstract}

\maketitle


\tableofcontents

\section{Introduction}

IR divergence is a significant issue in quantum field theory (QFT) \cite{Chung,Nakanishi}. While UV divergence can be dealt with using the technique of renormalization in renormalizable theories, IR divergence can not be dealt with in this way. The conventional way to deal with IR divergence for loop diagrams is to give the virtual photon $\gamma$ a positive mass $m_{\gamma}$. Cross sections and lifetimes can then be computed in the standard way but the final results depend on the parameter $m_{\gamma}$. 

IR divergences occur frequently for loop diagrams. However they also occur in tree level computations involving initial or final state radiation in which case the initial or final state photons may be given fictitious positive masses thereby enabling a finite result to be determined.

It has been found \cite{Jauch,Yennie,Bloch,Kinoshita,Lee,Akhoury} that when the cross sections for the IR divergent processes in QED involving loop diagrams are added to the cross sections associated with the final state (or initial state) radiation diagrams then the terms involving the fictitious photon mass precisely cancel.  This is formalized in the Bloch-Nordsieck theorem \cite{Bloch,Frye} for QED. In QCD cancellation may not occur at the 1-loop level but occurs when additional loops are considered \cite{Doria,Catani}. One has, in general for unitary theories, the Kinoshita-Lee-Nauenberg (KLN) theorem \cite{Kinoshita,Khalil}.

We have developed a method of regularization for QFT which we have called spectral regularization \cite{Springer,Symmetry,IJMPA,NPB}. We now call the technique covariant spectral regularization to distinguish it from other techniques called ``spectral regularization". In covariant spectral regularization UV or IR divergent integrals in QFT are interpreted as defining covariant measures and are analyzed using a spectral calculus leading to the determination of a density representing them with respect to Lebesgue measure on Minkowski space.

In Ref. \cite{NPB} we use the technique to compute the vacuum polarization tensor for QED and hence the Uehling contribution to the Lamb shift for the H atom. In Ref. \cite{Vertex_function} we use the technique to evaluate the vertex function in QED in the t channel and therefore derive Schwinger's result for the leading order (LO) contribution to the anomalous magnetic moment of the electron. We also compute the vertex function in the s channel which we use to compute the LO vertex correction contribution to the high energy limit of the cross section for the process $e^{+}e^{-}\rightarrow\mu^{+}\mu^{-}$ without encountering either UV or IR divergence. 

Our result in Ref. \cite{Vertex_function} agrees with the textbook result \cite{Schwartz} for the high energy limit for the LO contribution to the cross section for this process. In the textbook computation both UV and IR divergences are encountered. The UV divergence is dealt with by using Pauli-Villars or dimensional regularization together with renormalization while the IR divergence is cancelled by considering soft photon final state radiation. In our computation in Ref. \cite{Vertex_function} neither UV nor IR divergence is encountered so final state radiation does not need to be considered.

Since, in the standard approach, the tree level diagram for the process $e^{+}e^{-}\rightarrow\mu^{+}\mu^{-}\gamma$ is IR divergent and this divergence is used to cancel the IR divergence in the computation of the vertex function, we show, in the present paper, for consistency with the results of Ref.  \cite{Vertex_function}, that when analyzed using a careful treatment of distributional objects, the process $e^{+}e^{-}\rightarrow\mu^{+}\mu^{-}\gamma$ is not IR divergent at tree level in the soft photon high energy limit and that, in this limit, the cross section for the final state photons for this process vanishes.

In summary the results of this paper show that the so called ``infrared catastrophe" can be avoided by correct interpretation of a certain distributional object associated with the Feynman amplitude describing the process under consideration. 

In Section \ref{section:Feynman_amplitude} we compute the Feynman amplitude for the final state radiation process. In Section \ref{section:distributional_representation} we compute a meaningful distributional representation for a certain object arising from the Feynman amplitude for the process in the high energy limit. In Section \ref{section:Feyn_amp_sq} we compute the quantity $\overline{|{\mathcal M}|^2}$ where ${\mathcal M}$ is the Feynman amplitude for the process. In Section \ref{section:LIPS} we make some comments about the definition of the Lorentz invariant phase space and differential cross section for the process. In Section \ref{section:cross_section} we compute the cross section for the process in the soft photon high energy limit and establish the properties of it that we have mentioned above. The paper concludes with Section \ref{section:conclusion}. 
 
\section{Calculation of the Feynman amplitude for the process \label{section:Feynman_amplitude}}

\begin{figure} 
\centering
\includegraphics[width=6cm]{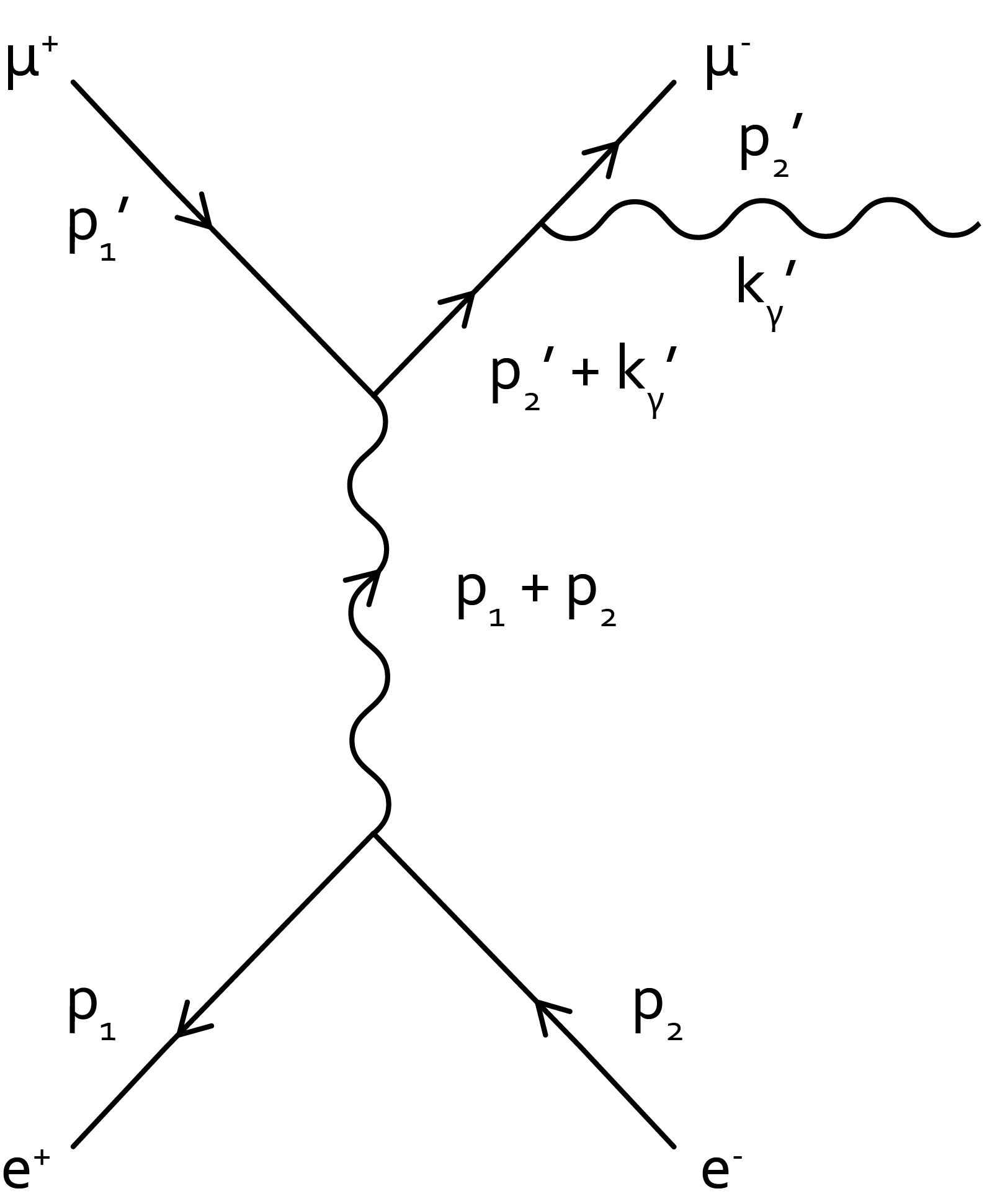}
\caption{First tree level Feynman diagram for process $e^{+}e^{-}\rightarrow\mu^{+}\mu^{-}\gamma$} \label{figure:diag_1}
\end{figure}

\begin{figure} 
\centering
\includegraphics[width=6cm]{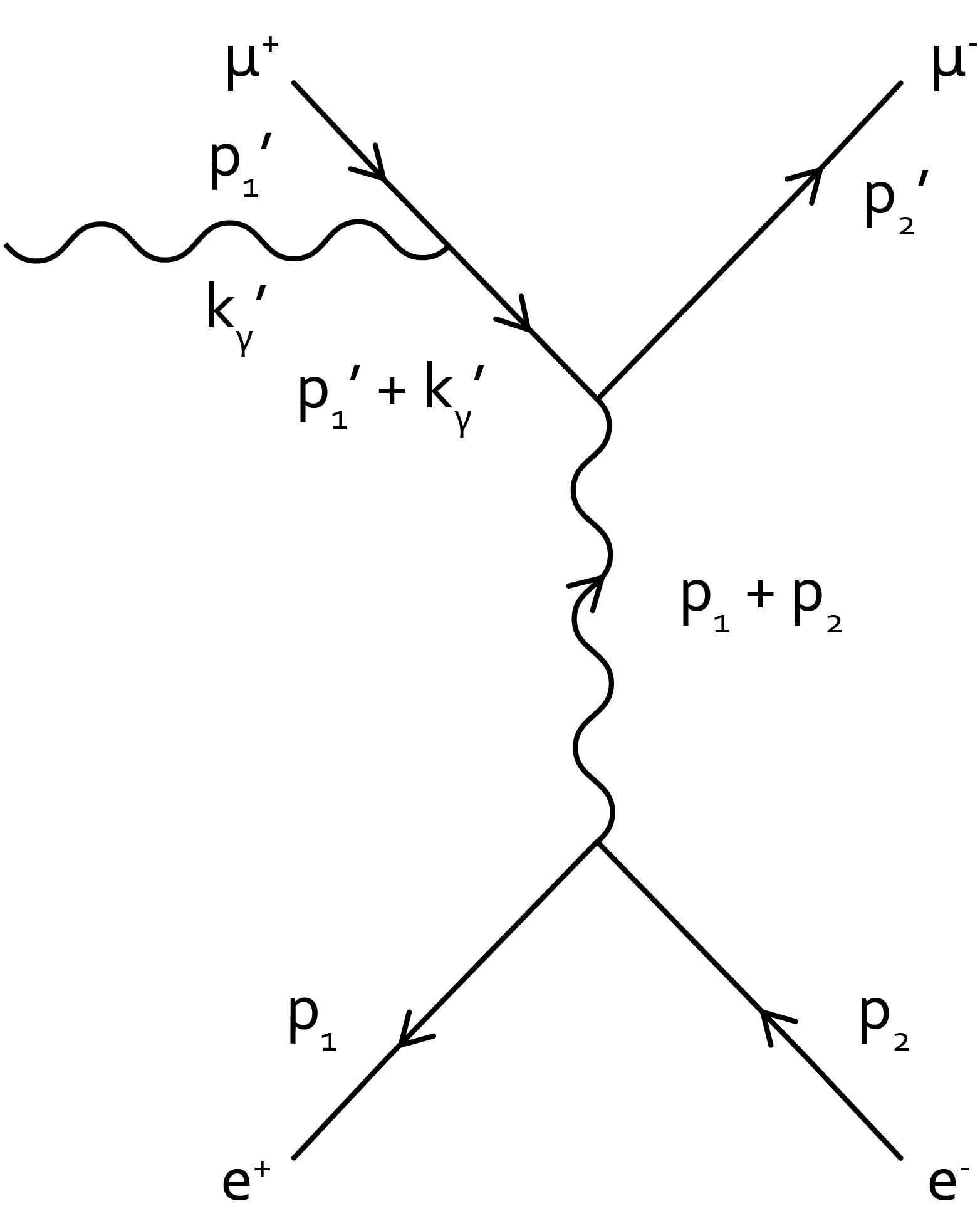}
\caption{Second tree level Feynman diagram for process $e^{+}e^{-}\rightarrow\mu^{+}\mu^{-}\gamma$} \label{figure:diag_2}
\end{figure}

The Feynman diagrams associated with the final state radiation process are shown in Figures \ref{figure:diag_1} and \ref{figure:diag_2}.

Using the Feynman rules we have that the Feynman amplitude for the process is given by
\begin{equation}
{\mathcal M}={\mathcal M}_1+{\mathcal M}_2,
\end{equation}
where
\begin{align*}
i{\mathcal M}_1=&\overline{v}_1(p_1,\alpha_1)ie\gamma^{\mu}u_2(p_2,\alpha_2)iD_{\mu\nu}(p_1+p_1)\overline{u}_2^{\prime}(p_2^{\prime},\alpha_2^{\prime})ie\gamma^{\rho^{\prime}}\epsilon^{\prime}(k_{\gamma}^{\prime},\rho^{\prime},\alpha^{\prime})^{*}\\
&iS(p_2^{\prime}+k_{\gamma}^{\prime},m_{\mu})ie\gamma^{\nu}v_1^{\prime}(p_1^{\prime},\alpha_1^{\prime}),\\
i{\mathcal M}_2=&\overline{v}_1(p_1,\alpha_1)ie\gamma^{\mu}u_2(p_2,\alpha_2)iD_{\mu\nu}(p_1+p_1)\overline{u}_2^{\prime}(p_2^{\prime},\alpha_2^{\prime})ie\gamma^{\nu}iS(p_1^{\prime}+k_{\gamma}^{\prime},m_{\mu})\\
&\epsilon^{\prime}(k_{\gamma}^{\prime}\sigma^{\prime},\alpha^{\prime})^{*}ie\gamma^{\sigma^{\prime}}v_1^{\prime}(p_1^{\prime},\alpha_1^{\prime}),
\end{align*}
with
\begin{equation}
D_{\mu\nu}(k)=\frac{-\eta_{\mu\nu}}{k^2+i\epsilon},
\end{equation}
the photon propagator and
\begin{equation}
S(p,m)=\frac{1}{{\slas p}-m+i\epsilon},
\end{equation}
the propagator for a fermion of mass $m$. 
Therefore
\begin{align*}
{\mathcal M}_{1\alpha_1^{\prime}\alpha_2^{\prime}\alpha_1\alpha_2}(p_1^{\prime},p_2^{\prime},p_1,p_2)=&-\frac{e^3}{Q^2}\overline{v}_1(p_1,\alpha_1)\gamma^{\mu}u_2(p_2,\alpha_2)\eta_{\mu\nu}\overline{u}_2^{\prime}(p_2^{\prime},\alpha_2^{\prime})\gamma^{\rho^{\prime}}\epsilon^{\prime}(k_{\gamma}^{\prime},\rho^{\prime},\alpha^{\prime})^{*}\\
&\frac{1}{{\slas p}_2^{\prime}+{\slas k}_{\gamma}^{\prime}-m_{\mu}+i\epsilon}\gamma^{\nu}v_1^{\prime}(p_1^{\prime},\alpha_1^{\prime}),\\
{\mathcal M}_{2\alpha_1^{\prime}\alpha_2^{\prime}\alpha_1\alpha_2}(p_1^{\prime},p_2^{\prime},p_1,p_2)=&-\frac{e^3}{Q^2}\overline{v}_1(p_1,\alpha_1)\gamma^{\mu}u_2(p_2,\alpha_2)\eta_{\mu\nu}\overline{u}_2^{\prime}(p_2^{\prime},\alpha_2^{\prime})\gamma^{\nu}\frac{1}{{\slas p}_1^{\prime}+{\slas k}_{\gamma}^{\prime}-m_{\mu}+i\epsilon}\\
&\epsilon^{\prime}(k_{\gamma^{\prime}},\rho^{\prime},\alpha^{\prime})^{*}\gamma^{\rho^{\prime}}v_1^{\prime}(p_1^{\prime},\alpha_1^{\prime}),
\end{align*}
where 
\begin{equation}
Q=((p_1+p_2)^2)^{\frac{1}{2}}.
\end{equation}
Thus
\begin{align*}
{\mathcal M}_1^{\dagger}=&-\frac{e^3}{Q^2}v_1^{\prime}(p_1^{\prime},\alpha_1^{\prime})^{\dagger}\gamma^0\gamma^{\nu}\gamma^0\left(\frac{1}{(p_2^{\prime}+k_{\gamma}^{\prime})^2-m_{\mu}^2+i\epsilon}\right)^{\dagger}(\gamma^0({\slas p}_2^{\prime}+{\slas k}_{\gamma}^{\prime}+m_{\mu})\gamma^0)\epsilon^{\prime}(k_{\gamma}^{\prime},\rho^{\prime},\alpha^{\prime})\\
&\gamma^0\gamma^{\rho^{\prime}}\gamma^0\gamma^0u_2^{\prime}(p_2^{\prime},\alpha_2^{\prime})\eta_{\mu\nu}u_2(p_2,\alpha_2)^{\dagger}\gamma^0\gamma^{\mu}\gamma^0\gamma^0v_1(p_1,\alpha_1)\\
=&-\frac{e^3}{Q^2}\overline{v}_1^{\prime}(p_1^{\prime},\alpha_1^{\prime})\gamma^{\nu}\left(\frac{1}{(p_2^{\prime}+k_{\gamma}^{\prime})^2-m_{\mu}^2+i\epsilon}\right)^{\dagger}({\slas p}_2^{\prime}+{\slas k}_{\gamma}^{\prime}+m_{\mu})\epsilon^{\prime}(k_{\gamma}^{\prime},\rho^{\prime},\alpha^{\prime})\gamma^{\rho^{\prime}}u_2^{\prime}(p_2^{\prime},\alpha_2^{\prime})\eta_{\mu\nu}\\
&\overline{u}_2(p_2,\alpha_2)\gamma^{\mu}v_1(p_1,\alpha_1).
\end{align*}
and
\begin{align*}
{\mathcal M}_2^{\dagger}=&-\frac{e^3}{Q^2}v_1^{\prime}(p_1^{\prime},\alpha_1^{\prime})^{\dagger}\gamma^0\gamma^{\rho^{\prime}}\gamma^0\epsilon^{\prime}(k_{\gamma}^{\prime},\rho^{\prime},\alpha^{\prime})\left(\frac{1}{(p_1^{\prime}+k_{\gamma}^{\prime})^2-m_{\mu}^2+i\epsilon}\right)^{\dagger}(\gamma^0({\slas p}_1^{\prime}+{\slas k}_{\gamma}^{\prime}+m_{\mu})\\
&\gamma^0)\gamma^0\gamma^{\nu}\gamma^0\gamma^0u_2^{\prime}(p_2^{\prime},\alpha_2^{\prime})\eta_{\mu\nu}u_2(p_2,\alpha_2)^{\dagger}\gamma^0\gamma^{\mu}\gamma^0\gamma^0v_1(p_1,\alpha_1)\\
=&-\frac{e^3}{Q^2}\overline{v}_1^{\prime}(p_1^{\prime},\alpha_1^{\prime})\gamma^{\rho^{\prime}}\epsilon^{\prime}(k_{\gamma}^{\prime},\rho^{\prime},\alpha^{\prime})\left(\frac{1}{(p_1^{\prime}+k_{\gamma}^{\prime})^2-m_{\mu}^2+i\epsilon}\right)^{\dagger}({\slas p}_1^{\prime}+{\slas k}_{\gamma}^{\prime}+m_{\mu})\gamma^{\nu}u_2^{\prime}(p_2^{\prime},\alpha_2^{\prime})\eta_{\mu\nu}\\
&\overline{u}_2(p_2,\alpha_2)\gamma^{\mu}v_1(p_1,\alpha_1).
\end{align*}
Therefore
\begin{align*}
\sum_{\mbox{spins,pols}}{\mathcal M}_1^{\dagger}{\mathcal M}_1=&\frac{e^6}{Q^4}\sum_{\alpha_1^{\prime},\alpha_2^{\prime},\alpha_1,\alpha_2,\alpha^{\prime}=1}^2\overline{v}_1^{\prime}(p_1^{\prime},\alpha_1^{\prime})\gamma^{\nu}\left(\frac{1}{(p_2^{\prime}+k_{\gamma}^{\prime})^2-m_{\mu}^2+i\epsilon}\right)^{\dagger}({\slas p}_2^{\prime}+{\slas k}_{\gamma}^{\prime}+m_{\mu})\\
&\epsilon^{\prime}(k_{\gamma}^{\prime},\sigma^{\prime},\alpha^{\prime})\gamma^{\sigma^{\prime}}u_2^{\prime}(p_2^{\prime},\alpha_2^{\prime})\eta_{\mu\nu}\overline{u}_2(p_2,\alpha_2)\gamma^{\mu}v_1(p_1,\alpha_1)\overline{v}_1(p_1,\alpha_1)\gamma^{\mu^{\prime}}u_2(p_2,\alpha_2)\\
&\eta_{\mu^{\prime}\nu^{\prime}}\overline{u}_2^{\prime}(p_2^{\prime},\alpha_2^{\prime})\gamma^{\rho^{\prime}}\epsilon^{\prime}(k_{\gamma}^{\prime},\rho^{\prime},\alpha^{\prime})^{*}\frac{1}{(p_2^{\prime}+k_{\gamma}^{\prime})^2-m_{\mu}^2+i\epsilon}({\slas p}_2^{\prime}+{\slas k}_{\gamma}^{\prime}+m_{\mu})\\
&\gamma^{\nu^{\prime}}v_1^{\prime}(p_1^{\prime},\alpha_1^{\prime})\\
=&\frac{e^6}{Q^4}\sum_{\alpha_1^{\prime},\alpha_2^{\prime},\alpha_1,\alpha_2,\alpha^{\prime}=1}^2\overline{v}_1^{\prime}(p_1^{\prime},\alpha_1^{\prime})_{\rho_1}\gamma^{\nu\rho_1}{}_{\rho_2}\left(\frac{1}{(p_2^{\prime}+k_{\gamma}^{\prime})^2-m_{\mu}^2+i\epsilon}\right)^{\dagger}\\
&({\slas p}_2^{\prime}+{\slas k}_{\gamma}^{\prime}+m_{\mu})^{\rho_2}{}_{\rho_3}\epsilon^{\prime}(k_{\gamma}^{\prime},\sigma^{\prime},\alpha^{\prime})\gamma^{\sigma^{\prime}\rho_3}{}_{\rho_4}u_2^{\prime}(p_2^{\prime},\alpha_2^{\prime})^{\rho_4}\eta_{\mu\nu}\overline{u}_2(p_2,\alpha_2)_{\rho_5}\gamma^{\mu\rho_5}{}_{\rho_6}\\
&v_1(p_1,\alpha_1)^{\rho_6}\overline{v}_1(p_1,\alpha_1)_{\sigma_1}\gamma^{\mu^{\prime}\sigma_1}{}_{\sigma_2}u_2(p_2,\alpha_2)^{\sigma_2}\eta_{\mu^{\prime}\nu^{\prime}}\overline{u}_2^{\prime}(p_2^{\prime},\alpha_2^{\prime})_{\sigma_3}\gamma^{\rho^{\prime}\sigma_3}{}_{\sigma_4}\\
&\epsilon^{\prime}(k_{\gamma}^{\prime},\rho^{\prime},\alpha^{\prime})^{*}\frac{1}{(p_2^{\prime}+k_{\gamma}^{\prime})^2-m_{\mu}^2+i\epsilon}({\slas p}_2^{\prime}+{\slas k}_{\gamma}^{\prime}+m_{\mu})^{\sigma_4}{}_{\sigma_5}\gamma^{\nu^{\prime}\sigma_5}{}_{\sigma_6}v_1^{\prime}(p_1^{\prime},\alpha_1^{\prime})^{\sigma_6}\\
=&\frac{e^6}{Q^4}({\slas p}_1^{\prime}-m_{\mu})^{\sigma_6}{}_{\rho_1}\gamma^{\nu\rho_1}{}_{\rho_2}\left(\frac{1}{(p_2^{\prime}+k_{\gamma}^{\prime})^2-m_{\mu}^2+i\epsilon}\right)^{\dagger}\\
&({\slas p}_2^{\prime}+{\slas k}_{\gamma}^{\prime}+m_{\mu})^{\rho_2}{}_{\rho_3}(-\eta_{\rho^{\prime}\sigma^{\prime}})\gamma^{\sigma^{\prime}\rho_3}{}_{\rho_4}({\slas p}_2^{\prime}+m_{\mu})^{\rho_4}{}_{\sigma_3}\eta_{\mu\nu}({\slas p}_2+m_e)^{\sigma_2}{}_{\rho_5}\gamma^{\mu\rho_5}{}_{\rho_6}\\
&({\slas p}_1-m_e)^{\rho_6}{}_{\sigma_1}\gamma^{\mu^{\prime}\sigma_1}{}_{\sigma_2}\eta_{\mu^{\prime}\nu^{\prime}}\gamma^{\rho^{\prime}\sigma_3}{}_{\sigma_4}\frac{1}{(p_2^{\prime}+k_{\gamma}^{\prime})^2-m_{\mu}^2+i\epsilon}\\
&({\slas p}_2^{\prime}+{\slas k}_{\gamma}^{\prime}+m_{\mu})^{\sigma_4}{}_{\sigma_5}\gamma^{\nu^{\prime}\sigma_5}{}_{\sigma_6}\\
=&-\frac{e^6}{Q^4}\eta_{\mu\nu}\eta_{\mu^{\prime}\nu^{\prime}}\eta_{\rho^{\prime}\sigma^{\prime}}\left|\frac{1}{(p_2^{\prime}+k_{\gamma}^{\prime})^2-m_{\mu}^2+i\epsilon}\right|^2\\
&\mbox{Tr}[({\slas p}_1^{\prime}-m_{\mu})\gamma^{\nu}({\slas p}_2^{\prime}+{\slas k}_{\gamma}^{\prime}+m_{\mu})\gamma^{\sigma^{\prime}}({\slas p}_2^{\prime}+m_{\mu})\gamma^{\rho^{\prime}}({\slas p}_2^{\prime}+{\slas k}_{\gamma}^{\prime}+m_{\mu})\gamma^{\nu^{\prime}}]\\
&\mbox{Tr}[({\slas p}_2+m_e)\gamma^{\mu}({\slas p}_1-m_e)\gamma^{\mu^{\prime}}]
\end{align*}
Since
\[ \eta_{\rho^{\prime}\sigma^{\prime}}\gamma^{\sigma^{\prime}}({\slas p}_2^{\prime}+m_{\mu})\gamma^{\rho^{\prime}}=-2{\slas p}_2^{\prime}+4m_{\mu}, \]
we have
\begin{align*}
\sum_{\mbox{spins,pols}}{\mathcal M}_1^{\dagger}{\mathcal M}_1=&-\frac{e^6}{Q^4}\eta_{\mu\nu}\eta_{\mu^{\prime}\nu^{\prime}}\left|\frac{1}{(p_2^{\prime}+k_{\gamma}^{\prime})^2-m_{\mu}^2+i\epsilon}\right|^2\\
&\mbox{Tr}[({\slas p}_1^{\prime}-m_{\mu})\gamma^{\nu}({\slas p}_2^{\prime}+{\slas k}_{\gamma}^{\prime}+m_{\mu})(-2{\slas p}_2^{\prime}+4m_{\mu})({\slas p}_2^{\prime}+{\slas k}_{\gamma}^{\prime}+m_{\mu})\gamma^{\nu^{\prime}}]\\
&\mbox{Tr}[({\slas p}_2+m_e)\gamma^{\mu}({\slas p}_1-m_e)\gamma^{\mu^{\prime}}].
\end{align*}
One can determine a distributional representation for the object $|(p_2^{\prime}+k_{\gamma}^{\prime})^2-m_{\mu}^2+i\epsilon|^{-2}$ as in Ref.  \cite{Vertex_function} and proceed to compute the cross section for the process under consideration including the distribution of energy of the final state radiation. However this is quite complicated and in the present paper we are principally concerned with the demonstration of the non-divergence of the process in the soft photon high energy limit and the computation of the cross section for the process in this limit.
Thus we will consider this limit where $m_e$ and $m_{\mu}$ can be neglected in comparison with $p_1,p_2,p_1^{\prime},p_2^{\prime},p_1^{\prime}+k_{\gamma}^{\prime}$ and $p_2^{\prime}+k_{\gamma}^{\prime}$. In this limit we have
\begin{align}\label{eq:fs_4}
\sum_{\mbox{spins,pols}}{\mathcal M}_1^{\dagger}{\mathcal M}_1=&2\frac{e^6}{Q^4}\eta_{\mu\nu}\eta_{\mu^{\prime}\nu^{\prime}}\left|\frac{1}{(p_2^{\prime}+k_{\gamma}^{\prime})^2-m_{\mu}^2+i\epsilon}\right|^2\Phi_1\Phi_2,
\end{align}
where
\begin{align*}
\Phi_1=&\mbox{Tr}[{\slas p}_1^{\prime}\gamma^{\nu}({\slas p}_2^{\prime}+{\slas k}_{\gamma}^{\prime}){\slas p}_2^{\prime}({\slas p}_2^{\prime}+{\slas k}_{\gamma}^{\prime})\gamma^{\nu^{\prime}}],\\
\Phi_2=&\mbox{Tr}[{\slas p}_2\gamma^{\mu}{\slas p}_1\gamma^{\mu^{\prime}}].
\end{align*}
We now compute (in the high energy limit)
\begin{align} \label{eq:miraculous_cancellation}
\Phi_1=&\mbox{Tr}[{\slas p}_1^{\prime}\gamma^{\nu}({\slas p}_2^{\prime}+{\slas k}_{\gamma}^{\prime}){\slas p}_2^{\prime}({\slas p}_2^{\prime}+{\slas k}_{\gamma}^{\prime})\gamma^{\nu^{\prime}}]\\
=&\mbox{Tr}[{\slas p}_1^{\prime}\gamma^{\nu}(p_{2\alpha}^{\prime}+k_{\gamma\alpha}^{\prime})p_{2\beta}^{\prime}\gamma^{\alpha}\gamma^{\beta}({\slas p}_2^{\prime}+{\slas k}_{\gamma}^{\prime})\gamma^{\nu^{\prime}}]\nonumber\\
=&\mbox{Tr}[{\slas p}_1^{\prime}\gamma^{\nu}(p_{2\alpha}^{\prime}+k_{\gamma\alpha}^{\prime})p_{2\beta}^{\prime}(2\eta_{\alpha\beta}-\gamma^{\beta}\gamma^{\alpha})({\slas p}_2^{\prime}+{\slas k}_{\gamma}^{\prime})\gamma^{\nu^{\prime}}]\nonumber\\
=&2\mbox{Tr}[{\slas p}_1^{\prime}\gamma^{\nu}((p_2^{\prime}+k_{\gamma}^{\prime}).p_2^{\prime})({\slas p}_2^{\prime}+{\slas k}_{\gamma}^{\prime})\gamma^{\nu^{\prime}}]-\mbox{Tr}[{\slas p}_1^{\prime}\gamma^{\nu}{\slas p}_2^{\prime}({\slas p}_2^{\prime}+{\slas k}_{\gamma}^{\prime})^2\gamma^{\nu^{\prime}}]\nonumber\\
=&2(m_{\mu}^2+(p_2^{\prime}.k_{\gamma}^{\prime}))\mbox{Tr}[{\slas p}_1^{\prime}\gamma^{\nu}({\slas p}_2^{\prime}+{\slas k}_{\gamma}^{\prime})\gamma^{\nu^{\prime}}]-\mbox{Tr}[{\slas p}_1^{\prime}\gamma^{\nu}{\slas p}_2^{\prime}(m_{\mu}^2+k_{\gamma}^{\prime2}+2p_2^{\prime}.k_{\gamma}^{\prime})\gamma^{\nu^{\prime}}]\nonumber\\
=&2(p_2^{\prime}.k_{\gamma}^{\prime})\mbox{Tr}[{\slas p}_1^{\prime}\gamma^{\nu}({\slas p}_2^{\prime}+{\slas k}_{\gamma}^{\prime})\gamma^{\nu^{\prime}}]-2(p_2.k_{\gamma}^{\prime})\mbox{Tr}[{\slas p}_1^{\prime}\gamma^{\nu}{\slas p}_2^{\prime}\gamma^{\nu^{\prime}}]\nonumber\\
=&2(p_2^{\prime}.k_{\gamma}^{\prime})\mbox{Tr}[{\slas p}_1^{\prime}\gamma^{\nu}{\slas k}_{\gamma}^{\prime}\gamma^{\nu^{\prime}}]\nonumber\\
=&2(p_2^{\prime}.k_{\gamma}^{\prime})p_{1\alpha}^{\prime}k_{\gamma\beta}^{\prime}\mbox{Tr}[\gamma^{\alpha}\gamma^{\nu}\gamma^{\beta}\gamma^{\nu^{\prime}}]\nonumber\\
=&8(p_2^{\prime}.k_{\gamma}^{\prime})p_{1\alpha}^{\prime}k_{\gamma\beta}^{\prime}(\eta^{\alpha\nu}\eta^{\beta\nu^{\prime}}-\eta^{\alpha\beta}\eta^{\nu\nu^{\prime}}+\eta^{\alpha\nu^{\prime}}\eta^{\nu\beta})\nonumber\\
=&8(p_2^{\prime}.k_{\gamma}^{\prime})(p_1^{\prime\nu}k_{\gamma}^{\prime\nu^{\prime}}-(p_1^{\prime}.k_{\gamma}^{\prime})\eta^{\nu\nu^{\prime}}+p_1^{\prime\nu^{\prime}}k_{\gamma}^{\prime\nu}).\nonumber
\end{align}
Note the cancellation that occurs between lines 6 and 7 of this computatiom. While it may be an exaggeration to describe it as a ``miraculous" cancellation one may notice that it is sufficient to ensure the infrared finiteness of the cross section for the process under consideration. Also
\begin{align*}
\Phi_2=&p_{2\alpha}p_{1\beta}\mbox{Tr}[\gamma^{\alpha}\gamma^{\mu}\gamma^{\beta}\gamma^{\mu^{\prime}}]\\
=&4p_{2\alpha}p_{1\beta}(\eta^{\alpha\mu}\eta^{\beta\mu^{\prime}}-\eta^{\alpha\beta}\eta^{\mu\mu^{\prime}}+\eta^{\alpha\mu^{\prime}}\eta^{\mu\beta})\\
=&4(p_2^{\mu}p_1^{\mu^{\prime}}-(p_2.p_1)\eta^{\mu\mu^{\prime}}+p_2^{\mu^{\prime}}p_1^{\mu}).
\end{align*}
Therefore
\begin{align}
\eta_{\mu\nu}\eta_{\mu^{\prime}\nu^{\prime}}\Phi_1\Phi_2=&\,32(p_2^{\prime}.k_{\gamma}^{\prime})(p_{1\mu}^{\prime}k_{\gamma\mu^{\prime}}^{\prime}-(p_1^{\prime}.k_{\gamma}^{\prime})\eta_{\mu\mu^{\prime}}+p_{1\mu^{\prime}}^{\prime}k_{\gamma\mu}^{\prime})\nonumber\\
&(p_2^{\mu}p_1^{\mu^{\prime}}-(p_2.p_1)\eta^{\mu\mu^{\prime}}+p_2^{\mu^{\prime}}p_1^{\mu})\nonumber\\
=&\,32(p_2^{\prime}.k_{\gamma}^{\prime})((p_1^{\prime}.p_2)(k_{\gamma}^{\prime}.p_1)-(p_2.p_1)(p_1^{\prime}.k_{\gamma}^{\prime})+(p_1^{\prime}.p_1)(p_2.k_{\gamma}^{\prime})-(p_1^{\prime}.k_{\gamma}^{\prime})(p_2.p_1)\nonumber\\
+&\,4(p_1^{\prime}.k_{\gamma}^{\prime})(p_2.p_1)-(p_1^{\prime}.k_{\gamma}^{\prime})(p_2.p_1)+(p_1.p_1^{\prime})(p_2.k_{\gamma}^{\prime})\nonumber\\
-&\,(p_2.p_1)(p_1^{\prime}.k_{\gamma}^{\prime})+(p_2.p_1^{\prime})(p_1.k_{\gamma}^{\prime})\nonumber\\
=&\,64(p_2^{\prime}.k_{\gamma}^{\prime})((p_2.p_1^{\prime})(p_1.k_{\gamma}^{\prime})+(p_1.p_1^{\prime})(p_2.k_{\gamma}^{\prime})).\label{eq:fs_5}
\end{align}
Since $p_2^{\prime}$ and $k_{\gamma}^{\prime}$ are on shell we have
\[ (p_2^{\prime}+k_{\gamma}^{\prime})^2-m_{\mu}^2=2p_2^{\prime}.k_{\gamma}^{\prime}. \]
Also
\[ p_2^{\prime}.k_{\gamma}^{\prime}=\omega_{m_{\mu}}({\vct p}_2^{\prime})|{\vct k}_{\gamma}^{\prime}|-{\vct p}_2^{\prime}.{\vct k}_{\gamma}^{\prime}>0, \]
for ${\vct k}_{\gamma}^{\prime}\neq0$, where, for any $m\geq0,{\vct p}\in{\bf R}^3$,
\begin{equation}
\omega_m({\vct p})=(m^2+{\vct p}^2)^{\frac{1}{2}}.
\end{equation}
Thus
\begin{align}
(p_2^{\prime}.k_{\gamma}^{\prime})\left|\frac{1}{(p_2^{\prime}+k_{\gamma}^{\prime})^2-m_{\mu}^2+i\epsilon}\right|^2=&(p_2^{\prime}.k_{\gamma}^{\prime})\left|\left(\frac{1}{2(p_2^{\prime}.k_{\gamma}^{\prime})+i\epsilon}\right)^2\right|=\frac{1}{4}\left|\frac{(p_2^{\prime}.k_{\gamma}^{\prime})}{((p_2^{\prime}.k_{\gamma}^{\prime})+i\epsilon)^2}\right|\nonumber\\ 
=&\frac{1}{4}\left|\frac{1}{(p_2^{\prime}.k_{\gamma}^{\prime})+i\epsilon}\right|.\label{eq:fs_6}
\end{align}
From Eqns.~\ref{eq:fs_4},\ref{eq:fs_5} and \ref{eq:fs_6}
\begin{align*}
\frac{1}{4}\sum_{\mbox{spins,pols}}{\mathcal M}_1^{\dagger}{\mathcal M}_1=&8\frac{e^6}{Q^4}\left|\frac{1}{(p_2^{\prime}.k_{\gamma}^{\prime})+i\epsilon}\right|((p_2.p_1^{\prime})(p_1.k_{\gamma}^{\prime})+(p_1.p_1^{\prime})(p_2.k_{\gamma}^{\prime})).
\end{align*}

\section{Determination of a distributional representation for the \mbox{ \,\, \, \, \, \, \, \, \, \, } object $\bf{((p_2^{\prime}.k_{\gamma}^{\prime})+i\epsilon)^{-1}}$ \label{section:distributional_representation}}

Now fix $p_2^{\prime}$ and we will now evaluate $f(k_{\gamma}^{\prime})=\frac{1}{(p_2^{\prime}.k_{\gamma}^{\prime})+i\epsilon}$. To do so we will compute its inverse Fourier transform, which is given by
\begin{equation}
{\ichp f}(x)=(2\pi)^{-4}\int\frac{1}{(p_2^{\prime}.k_{\gamma}^{\prime})+i\epsilon}e^{ik_{\gamma}^{\prime}.x}\,dk_{\gamma}^{\prime}.
\end{equation} 
We have that
\[ p_2^{\prime}.k_{\gamma}^{\prime}=p_2^{\prime0}k_{\gamma}^{\prime0}-{\vct p}_2^{\prime}.{\vct k}_{\gamma}^{\prime}, \]
where $p_2^{\prime0}=\omega_{m_{\mu}}({\vct p}_2^{\prime}),k_{\gamma}^{\prime0}=|{\vct k}_{\gamma}^{\prime}|$. Therefore
\begin{align*}
{\ichp f}(x)=&(2\pi)^{-4}(p_2^{\prime0})^{-1}\int\frac{1}{k_{\gamma}^{\prime0}-a({\vct p}_2^{\prime},{\vct k}_{\gamma}^{\prime})+i\epsilon}e^{ik_{\gamma}^{\prime}.x}\,dk_{\gamma}^{\prime}\\
=&(2\pi)^{-4}(p_2^{\prime0})^{-1}\int_{{\vct k}_{\gamma}^{\prime}\in{\bf R}^3}(\int_{k_{\gamma}^{\prime0}=-\infty}^{\infty}\frac{1}{k_{\gamma}^{\prime0}-a({\vct p}_2^{\prime},{\vct k}_{\gamma}^{\prime})+i\epsilon}e^{ik_{\gamma}^{\prime0}x^0}\,dk_{\gamma}^{\prime0})e^{-i{\vct k}_{\gamma}^{\prime}.{\vct x}}\,d{\vct k}_{\gamma}^{\prime},
\end{align*} 
where 
\begin{equation}
a({\vct p}_2^{\prime},{\vct k}_{\gamma}^{\prime})=(p_2^{\prime0})^{-1}({\vct p}_2^{\prime}.{\vct k}_{\gamma}^{\prime}).
\end{equation}
We want to compute, for the inner integral, a contour integral of the form
\[ I(a,c)=\int_{C_a}\frac{1}{z-a}e^{icz}\,dz, \]
where $a,c>0$ and
\begin{align*}
&C_a=C_{a1}\cup C_{a2}\cup C_{a3},\\
&C_{a1}(t)=t,t\in(-\infty,a-\epsilon),C_{a2}(t)=a+\epsilon e^{it},t\in[0,\pi],C_{a3}(t)=t,t\in(a+\epsilon,\infty).
\end{align*}
It is easy to show that
\begin{equation}
I(a,c)=\int_{C_0}\frac{1}{z}e^{ic(z+a)}\,dz=e^{ica}\int_{C_0}\frac{1}{z}e^{icz}\,dz.
\end{equation}
Now
\begin{align*}
\int_{C_{01}}\frac{1}{z}e^{icz}\,dz+\int_{C_{03}}\frac{1}{z}e^{icz}\,dz=&\int_{cC_{01}}\frac{1}{z}e^{iz}\,dz+\int_{cC_{03}}\frac{1}{z}e^{iz}\,dz\\
&\int_{t=-\infty}^{-c\epsilon}\frac{1}{t}(\cos(t)+i\sin(t))\,dt+\int_{t=c\epsilon}^{\infty}\frac{1}{t}(\cos(t)+i\sin(t))\,dt\\
=&2i\int_{c\epsilon}^{\infty}\mbox{sinc}(t)\,dt\rightarrow\pi i\mbox{ as }\epsilon\rightarrow0.
\end{align*}
Also
\[ \int_{C_{02}}\frac{1}{z}e^{icz}\,dz=\int_{t=0}^{\pi}\frac{1}{\epsilon e^{it}}e^{ic\epsilon e^{it}}(i\epsilon e^{it})\,dt=i\int_{t=0}^{\pi}e^{ic\epsilon e^{it}}\,dt, \]
and it is straightforward to show, using dominated convergence, that
\[ \int_{t=0}^{\pi}e^{ic\epsilon e^{it}}\,dt\rightarrow\int_{t=0}^{\pi}1\,dt=\pi\mbox{ as }\epsilon\rightarrow0. \]
Hence
\begin{equation}
I(a,c) (=I(a,c,\epsilon))\rightarrow2\pi i e^{ica}\mbox{ as }\epsilon\rightarrow0.
\end{equation}
Therefore
\begin{align*}
{\ichp f}(x)=&\,(2\pi)^{-4}(p_2^{\prime0})^{-1}\int_{{\vct k}_{\gamma}^{\prime}\in{\bf R}^3}(2\pi i)e^{i(a({\vct p}_2^{\prime},{\vct k}_{\gamma}^{\prime})x^0-{\vct k}_{\gamma}^{\prime}.{\vct x})}\,d{\vct k}_{\gamma}^{\prime}\\
=&\,(2\pi)^{-3}i(p_2^{\prime0})^{-1}\int_{{\vct k}_{\gamma}^{\prime}\in{\bf R}^3}e^{i((p_2^{\prime0})^{-1}x^0{\vct p}_2^{\prime}-{\vct x}).{\vct k}_{\gamma}^{\prime}}\,d{\vct k}_{\gamma}^{\prime}\\
=&\,i(p_2^{\prime0})^{-1}\delta({\vct x}-x^0(p_2^{\prime0})^{-1}{\vct p}_2^{\prime}).
\end{align*}
Thus the function (distribution) $f$ is given by the Fourier transform of ${\ichp f}$, i.e. 
\begin{align*}
f(k_{\gamma}^{\prime})=&\,i(p_2^{\prime0})^{-1}\int\delta({\vct x}-x^0(p_2^{\prime0})^{-1}{\vct p}_2^{\prime})e^{-ik_{\gamma}^{\prime}.x}\,dx\\
=&\,i(p_2^{\prime0})^{-1}\int_{x^0=-\infty}^{\infty}e^{-ik_{\gamma}^{\prime0}x^0}(\int_{{\vct x}\in{\bf R}^3}\delta({\vct x}-x^0(p_2^{\prime0})^{-1}{\vct p}_2^{\prime})e^{i{\vct k}_{\gamma}^{\prime}.{\vct x}}\,d{\vct x})\,dx^0\\
=&\,i(p_2^{\prime0})^{-1}\int_{x^0=-{\infty}}^{\infty}e^{-ik_{\gamma}^{\prime0}x^0+i{\vct k}_{\gamma}^{\prime}.(x^0(p_2^{\prime0})^{-1}{\vct p}_2^{\prime})}\,dx^0\\
=&\,i(p_2^{\prime0})^{-1}\int_{x^0=-{\infty}}^{\infty}e^{-i(k_{\gamma}^{\prime0}-(p_2^{\prime0})^{-1}({\vct p}_2^{\prime}.{\vct k}_{\gamma}^{\prime}))x^0}\,dx^0\\
=&\,2\pi i(p_2^{\prime0})^{-1}\delta(k_{\gamma}^{\prime0}-(p_2^{\prime0})^{-1}({\vct p}_2^{\prime}.{\vct k}_{\gamma}^{\prime})).
\end{align*}
Now
\[ k_{\gamma}^{\prime0}-(p_2^{\prime0})^{-1}({\vct p}_2^{\prime}.{\vct k}_{\gamma}^{\prime})=k_{\gamma}^{\prime0}(p_2^{\prime0})^{-1}(p_2^{\prime0}-|{\vct p}_2^{\prime}|({\vct\omega}_2^{\prime}.{\vct\omega_{\gamma}^{\prime}})), \]
where, for ${\vct p}_2\neq0$, ${\vct\omega}_2$ is defined by ${\vct\omega}_2=|{\vct p}_2|^{-1}{\vct p}_2$, similarly for ${\vct\omega}_{\gamma}^{\prime}$.
Furthermore
\[ |p_2^{\prime0}-|{\vct p}_2^{\prime}|({\vct\omega}_2^{\prime}.{\vct\omega}_{\gamma}^{\prime})|\geq p_2^{\prime0}-|{\vct p}_2^{\prime}||{\vct\omega}_2^{\prime}.{\vct\omega}_{\gamma}^{\prime}|\geq \omega_{m_{\mu}}({\vct p}_2)-|{\vct p}_2|>0. \]
Thus, using the relation $\delta(ax)=a^{-1}\delta(x)$, for $a>0$, and $x$ a real variable, we have
\[ (p_2^{\prime0})^{-1}\delta(k_{\gamma}^{\prime0}-(p_2^{\prime0})^{-1}({\vct p}_2^{\prime}.{\vct k}_{\gamma}^{\prime}))=\frac{1}{p_2^{\prime0}-|{\vct p}_2^{\prime}|({\vct\omega}_2^{\prime}.{\vct\omega}_{\gamma}^{\prime})}\delta(k_{\gamma}^{\prime0}). \]
Therefore
\begin{equation}
f(k_{\gamma}^{\prime})=2\pi i\frac{1}{p_2^{\prime0}-|{\vct p}_2^{\prime}|({\vct\omega}_2^{\prime}.{\vct\omega}_{\gamma}^{\prime})}\delta(k_{\gamma}^{\prime0}).
\end{equation}

\section{Computation of $\overline{|{\mathcal M}|^2}$\label{section:Feyn_amp_sq}}

From the computations of the previous sections we have that
\begin{align}
\frac{1}{4}\sum_{\mbox{spins,pols}}|{\mathcal M}_1|^2=&16\pi\frac{e^6}{Q^4}\frac{1}{p_2^{\prime0}-|{\vct p}_2^{\prime}|({\vct\omega}_2^{\prime}.{\vct\omega}_{\gamma}^{\prime})}\delta(k_{\gamma}^{\prime0})\nonumber\\
&((p_2.p_1^{\prime})(p_1.k_{\gamma}^{\prime})+(p_1.p_1^{\prime})(p_2.k_{\gamma}^{\prime})).
\end{align}
By a similar calculation it is straightforward to show that
\begin{align}
\frac{1}{4}\sum_{\mbox{spins,pols}}|{\mathcal M}_2|^2=&16\pi\frac{e^6}{Q^4}\frac{1}{p_1^{\prime0}-|{\vct p}_1^{\prime}|({\vct\omega}_1^{\prime}.{\vct\omega}_{\gamma}^{\prime})}\delta(k_{\gamma}^{\prime0})\nonumber\\
&((p_1.p_2^{\prime})(p_2.k_{\gamma}^{\prime})+(p_2.p_2^{\prime})(p_1.k_{\gamma}^{\prime})).
\end{align}
Also one can compute that
\begin{align*}
\frac{1}{4}\sum_{\mbox{spins,pols}}{\mathcal M}_1^{\dagger}{\mathcal M}_2=&16\pi i\frac{e^6}{Q^4}(p_2.p_1)(p_2^{\prime}.p_1^{\prime})\delta(k^{\prime0}-(p_2^{\prime0})^{-1}({\vct p}_2^{\prime}.{\vct k}_{\gamma}^{\prime})),
\end{align*}
which is pure imaginary and so 
\begin{equation}
\frac{1}{4}\sum_{\mbox{spins,pols}}({\mathcal M}_1^{\dagger}{\mathcal M}_2+{\mathcal M}_2^{\dagger}{\mathcal M}_1)=0.
\end{equation}
Putting together all the above computations we see that
\begin{align*}
\overline{|{\mathcal M}|^2}=&16\pi\frac{e^6}{Q^4}\delta(k_{\gamma}^{\prime0})\left\{\frac{1}{p_2^{\prime0}-|{\vct p}_2^{\prime}|({\vct\omega}_2^{\prime}.{\vct\omega}_{\gamma}^{\prime})}((p_2.p_1^{\prime})(p_1.k_{\gamma}^{\prime})+(p_1.p_1^{\prime})(p_2.k_{\gamma}^{\prime}))+\right.\\
&\left.\frac{1}{p_1^{\prime0}-|{\vct p}_1^{\prime}|({\vct\omega}_1^{\prime}.{\vct\omega}_{\gamma}^{\prime})}((p_1.p_2^{\prime})(p_2.k_{\gamma}^{\prime})+(p_2.p_2^{\prime})(p_1.k_{\gamma}^{\prime}))\right\}.
\end{align*}

\section{Lorentz invariant phase space and differential cross section for the process \label{section:LIPS}}

We consider the LIPS measure for $\gamma\rightarrow\mu^{+}\mu^{-}\gamma$. From the definition \cite{Schwartz} this measure is given by
\begin{align}
d\Pi_{\mbox{LIPS}}=&(2\pi)^4\delta(k-p_1^{\prime}-p_2^{\prime}-k_{\gamma}^{\prime})\frac{d{\vct p}_1^{\prime}}{(2\pi)^3}\frac{1}{2E_1^{\prime}}\frac{d{\vct p}_2^{\prime}}{(2\pi)^3}\frac{1}{2E_2^{\prime}}\frac{d{\vct k}_{\gamma}^{\prime}}{(2\pi)^3}\frac{1}{2E_{\gamma}^{\prime}},
\end{align}
where $k=p_1+p_2$ is the momentum of the virtual $\gamma$, $p_1^{\prime}$ and $p_2^{\prime}$ are the momenta of the outgoing muons, $k_{\gamma}^{\prime}$ is the momentum of the outgoing $\gamma$, $E_1^{\prime}=E_1^{\prime}({\vct p}_1^{\prime})=\omega_{m_{\mu}}({\vct p}_1^{\prime}),E_2^{\prime}=E_2^{\prime}({\vct p}_2^{\prime})=\omega_{m_{\mu}}({\vct p}_2^{\prime}),E_{\gamma}^{\prime}=E_{\gamma}^{\prime}({\vct k}_{\gamma}^{\prime})=|{\vct k}_{\gamma}^{\prime}|$.

Thus we may write
\begin{align*}
d\Pi_{\mbox{LIPS}}=&(2\pi)^4\delta(k-p_1^{\prime}-p_2^{\prime}-k_{\gamma}^{\prime})(2\pi)^{-3}\delta((p_1^{\prime})^2-m^2)(2\pi)^{-3}\delta((p_2^{\prime})^2-m^2)\\
&(2\pi)^{-3}\delta((k_{\gamma}^{\prime})^2)\,dp_1^{\prime}\,dp_2^{\prime}\,dk_{\gamma}^{\prime}.
\end{align*}
It is straightforward to show that $d\Pi_{\mbox{LIPS}}$ is Lorentz invariant in the sense that
\begin{align*}
&\int(\Lambda\psi)(k,p_1^{\prime},p_2^{\prime},k_{\gamma}^{\prime})\,d\Pi_{\mbox{LIPS}}(k,p_1^{\prime},p_2^{\prime},k_{\gamma}^{\prime})=\int\psi(k,p_1^{\prime},p_2^{\prime},k_{\gamma}^{\prime})\,d\Pi_{\mbox{LIPS}}(k,p_1^{\prime},p_2^{\prime},k_{\gamma}^{\prime}),
\end{align*}
for all $\psi\in{\mathcal S}(({\bf R}^4)^4,{\bf C}),\Lambda\in O(1,3)^{\uparrow+}$.

$d\Pi_{\mbox{LIPS}}$ is defined as a product of measures, which might seem problematic. However $d\Pi_{\mbox{LIPS}}$ can be given a rigorous definition as a Lorentz invariant measure on the space 
\begin{align}
X=&\{(p_1,p_2,p_1^{\prime},p_2^{\prime},k_{\gamma}^{\prime}):(p_1)^2=m_e^2,(p_2)^2=m_e^2,(p_1^{\prime})^2=m_{\mu}^2,(p_2^{\prime})^2=m_{\mu}^2,\\
&(k_{\gamma}^{\prime})^2=0,p_1+p_2=p_1^{\prime}+p_2^{\prime}+k_{\gamma}^{\prime}\}, \nonumber
\end{align}
of on shell momentum 5-tuples for which conservation of momentum is satisfied.

For soft photon final state radiation the LIPS measure can be written as
\begin{equation}
d\Pi_{\mbox{LIPS}}=d\Pi_{\mbox{LIPS},0}\frac{d{\vct k}_{\gamma}^{\prime}}{(2\pi)^3}\frac{1}{2E_{\gamma}^{\prime}},
\end{equation}
where $E_{\gamma}^{\prime}=|{\vct k}_{\gamma}^{\prime}|$, and
\begin{align}
d\Pi_{\mbox{LIPS},0}=&(2\pi)^4\delta(k-p_1^{\prime}-p_2^{\prime})\frac{d{\vct p}_1^{\prime}}{(2\pi)^3}\frac{1}{2E_1^{\prime}}\frac{d{\vct p}_2^{\prime}}{(2\pi)^3}\frac{1}{2E_2^{\prime}},
\end{align}
is the LIPS for the tree level process without final state radiation. 
Thus
\begin{equation}
d\Pi_{\mbox{LIPS}}=\frac{1}{16\pi^3}d\Pi_{\mbox{LIPS},0}\,d\Omega({\vct\omega}_{\gamma}^{\prime}),
\end{equation}
where $d\Omega:{\mathcal B}(S^2)\rightarrow[0,4\pi]$ is the area measure for the sphere $S^2$.

Applying the fundamental principle of QFT for the computation of differential cross sections (Schwartz, 2014 \cite{Schwartz}, p. 61-63) we have that, in the CM frame, the differential cross section for the process without final state radiation is given by
\begin{align*}
d\sigma_0=\frac{1}{(2E_1)(2E_2)}\left(|{\vct p}_1|\frac{E_{CM}}{E_1E_2}\right)^{-1}\overline{|{\mathcal M}_0|^2}\,d\Pi_{\mbox{LIPS},0}.
\end{align*}
Therefore, the soft photon final state radiation differential cross section is
\begin{align*}
d\sigma=\frac{1}{16\pi^3}\frac{1}{(2E_1)(2E_2)}\left(|{\vct p}_1|\frac{E_{CM}}{E_1E_2}\right)^{-1}\overline{|{\mathcal M}|^2}\,d\Pi_{\mbox{LIPS},0}\,d\Omega({\vct\omega}_{\gamma}^{\prime}),
\end{align*}
As is well known \cite{Schwartz} one can compute that, in the high energy limit, and CM frame,
\begin{equation}
\frac{1}{(2E_1)(2E_2)}\left(|{\vct p}_1|\frac{E_{CM}}{E_1E_2}\right)^{-1}\,d\Pi_{\mbox{LIPS},0}=\frac{1}{64\pi^2E_{CM}^2}\,d\Omega({\vct\omega}^{\prime}).
\end{equation}
Therefore, for the final state radiation process
\begin{equation}
d\sigma=\frac{1}{16\pi^3}\frac{1}{64\pi^2E_{CM}^2}\overline{|{\mathcal M}|^2}\,d\Omega({\vct\omega}^{\prime})\,\,d\Omega({\vct\omega}_{\gamma}^{\prime}).
\end{equation}
i.e.
\begin{equation}
\left(\frac{d\sigma}{d\Omega}\right)_{CM}=\frac{1}{16\pi^3}\frac{1}{64\pi^2E_{CM}^2}\overline{|{\mathcal M}|^2}.
\end{equation}

\section{Proof that the soft photon high energy limit of the final state photon cross section is zero\label{section:cross_section}}

We have that
\begin{equation}
\overline{|{\mathcal M}|^2}=F(p_1^{\prime},p_2^{\prime},k_{\gamma}^{\prime},p_1,p_2)E_{\gamma}^{\prime}\delta(E_{\gamma}^{\prime}),
\end{equation}
where, for $E_{\gamma}^{\prime}=|{\vct k}_{\gamma}^{\prime}|\neq0$,
\begin{align} 
F(p_1^{\prime},p_2^{\prime},k_{\gamma}^{\prime},p_1,p_2)=&16\pi\frac{e^6}{Q^4}\left\{(\frac{1}{p_2^{\prime0}-|{\vct p}_2^{\prime}|({\vct\omega}_2^{\prime}.{\vct\omega}_{\gamma}^{\prime})}\frac{1}{E_{\gamma}^{\prime}}((p_2.p_1^{\prime})(p_1.k_{\gamma}^{\prime})+\right.\\
&\left.\frac{}{}(p_1.p_1^{\prime})(p_2.k_{\gamma}^{\prime})))+(1\leftrightarrow2)\right\}. \label{eq:quotient}\nonumber
\end{align}
Consider the first term
\begin{align*}
16\pi\frac{e^6}{Q^4}\frac{1}{p_2^{\prime0}-|{\vct p}_2^{\prime}|({\vct\omega}_2^{\prime}.{\vct\omega}_{\gamma}^{\prime})}\frac{1}{E_{\gamma}^{\prime}}((p_2.p_1^{\prime})(p_1.k_{\gamma}^{\prime})+(p_1.p_1^{\prime})(p_2.k_{\gamma}^{\prime})).
\end{align*}
We have
\begin{align*}
\frac{1}{E_{\gamma}^{\prime}}((p_2.p_1^{\prime})(p_1.k_{\gamma}^{\prime})+(p_1.p_1^{\prime})(p_2.k_{\gamma}^{\prime}))=(p_2.p_1^{\prime})(p_1^0-({\vct p}_1.{\vct\omega}_{\gamma}^{\prime}))+(p_1.p_1^{\prime})(p_2^0-({\vct p}_2.{\vct\omega}_{\gamma}^{\prime})),
\end{align*}
and can compute
\begin{align*}
&\int_{{\vct\omega}_{\gamma}^{\prime}\in S^2}\left|\frac{p_1^0-({\vct p}_1.{\vct\omega}_{\gamma}^{\prime})}{p_2^{\prime0}-|{\vct p}_2^{\prime}|({\vct\omega}_2^{\prime}.{\vct\omega}_{\gamma}^{\prime})}\right|\,d\Omega({\vct\omega}_{\gamma}^{\prime})\\
\leq&2\pi\int_0^{\pi}\frac{2p_1^0}{p_2^{\prime0}-|{\vct p}_2^{\prime}|\cos(\theta)}\sin(\theta)\,d\theta\\
=&4\pi\frac{p_1^0}{|{\vct p}_2^{\prime}|}\log\left(\frac{p_2^{\prime0}+|{\vct p}_2^{\prime}|}{p_2^{\prime0}-|{\vct p}_2^{\prime}|}\right)\\
=&4\pi\frac{p_1^0}{|{\vct p}_2^{\prime}|}\log\left(\frac{(p_2^{\prime0}+|{\vct p}_2^{\prime}|)^2}{m_{\mu}^2}\right)\\
\leq&8\pi\frac{p_1^0}{|{\vct p}_2^{\prime}|}\log(p_2^{\prime0}+|{\vct p}_2^{\prime}|)\\
\leq&8\pi\frac{p_1^0}{|{\vct p}_2^{\prime}|}\log(2p_2^{\prime0})\\
\leq&8\pi\log(Q),
\end{align*}
in the high energy limit. 
Also
\begin{align*}
|p_2.p_1^{\prime}|=&|p_2^{0}p_1^{\prime0}-{\vct p}_2.{\vct p}_1^{\prime}|\\
\leq&p_2^0p_1^{\prime0}\\
\leq&Q^2.
\end{align*}
Therefore
\begin{align*}
\int_{{\vct\omega}_{\gamma}\in S^2}\left|\frac{1}{p_2^{\prime0}-|{\vct p}_2^{\prime}|({\vct\omega}_2.{\vct\omega}_{\gamma})}\frac{1}{E_{\gamma}^{\prime}}(p_2.p_1^{\prime})(p_1.k_{\gamma}^{\prime})\right|\,d\Omega({\vct\omega}_{\gamma}^{\prime})\leq8\pi Q^2\log(Q).
\end{align*}
Similarly
\begin{align*}
\int_{{\vct\omega}_{\gamma}\in S^2}\left|\frac{1}{p_2^{\prime0}-|{\vct p}_2^{\prime}|({\vct\omega}_2.{\vct\omega}_{\gamma})}\frac{1}{E_{\gamma}^{\prime}}(p_1.p_1^{\prime})(p_2.k_{\gamma}^{\prime})\right|\,d\Omega({\vct\omega}_{\gamma}^{\prime})\leq8\pi Q^2\log(Q).
\end{align*}
Now in the CM frame ${\vct\omega}_2^{\prime}=-{\vct\omega}_1^{\prime}={\vct\omega}^{\prime}$, say. Therefore
\begin{align*}
&\int_{{\vct\omega}^{\prime}\in S^2}\int_{{\vct\omega}_{\gamma}^{\prime}\in S^2}\left|\frac{}{}\right.\frac{1}{p_2^{\prime0}-|{\vct p}_2^{\prime}|({\vct\omega}_2^{\prime}.{\vct\omega}_{\gamma}^{\prime})}[(p_2.p_1^{\prime})(p_1^0-({\vct p}_1.{\vct\omega}_{\gamma}^{\prime}))+\\
&\left.\frac{}{}(p_1.p_1^{\prime})(p_2^0-({\vct p}_2.{\vct\omega}_{\gamma}^{\prime}))]\right|\,d\Omega({\vct\omega}^{\prime})\,d\Omega({\vct\omega}_{\gamma}^{\prime})\\
\leq&\int_{{\vct\omega}^{\prime}\in S^2}\int_{{\vct\omega}_{\gamma}^{\prime}\in S^2}(16\pi Q^2\log(Q))\,d\Omega({\vct\omega}^{\prime})\,d\Omega({\vct\omega}_{\gamma}^{\prime})\\
\leq&(4\pi)^2(16\pi Q^2\log(Q))\\
<&\infty.
\end{align*}
Similarly for the second term of Eq. \eqref{eq:quotient}.
Therefore the integral
\begin{align*}
I(Q)=&\int_{{\vct\omega}^{\prime}\in S^2}\int_{{\vct\omega}_{\gamma}^{\prime}\in S^2}\left\{(\frac{1}{p_2^{\prime0}-|{\vct p}_2^{\prime}|({\vct\omega}_2^{\prime}.{\vct\omega}_{\gamma}^{\prime})}((p_2.p_1^{\prime})(p_1.k_{\gamma}^{\prime})+\right.\\
&\left.\frac{}{}(p_1.p_1^{\prime})(p_2.k_{\gamma}^{\prime})))+(1\leftrightarrow2)\right\}\,d\Omega({\vct\omega}^{\prime})\,d\Omega({\vct\omega}_{\gamma}^{\prime}),
\end{align*}
exists and is finite for all $Q>0$.

Therefore the cross section for the soft photon high energy limit for the final state radiation process can be written as
\begin{align*}
\sigma=&\sigma(Q,E_{\gamma}^{\prime})\\
=&\int_{{\vct\omega}^{\prime}\in S^2}\int_{{\vct\omega}_{\gamma}^{\prime}\in S^2}\left(\frac{d\sigma}{d\Omega}\right)_{CM}\,d\Omega({\vct\omega}^{\prime})\,d\Omega({\vct\omega}_{\gamma}^{\prime})\\
=&\frac{1}{16\pi^3}\frac{1}{64\pi^2E_{CM}^2}\int_{{\vct\omega}^{\prime}\in S^2}\int_{{\vct\omega}_{\gamma}^{\prime}\in S^2}\overline{|{\mathcal M}|^2}\,d\Omega({\vct\omega}^{\prime})\,d\Omega({\vct\omega}_{\gamma}^{\prime})\\
=&\frac{1}{16\pi^3}\frac{1}{64\pi^2E_{CM}^2}16\pi\frac{e^6}{Q^4}\delta(E_{\gamma}^{\prime})E_{\gamma}^{\prime}I(Q),
\end{align*}
and we have
\begin{align}
\lim_{E_{\gamma,\mbox{\scriptsize{soft}}}^{\prime}\rightarrow0}\int_0^{E_{\gamma\mbox{\scriptsize{soft}}}^{\prime}}\sigma(Q,E_{\gamma}^{\prime})\,dE_{\gamma}^{\prime}=0.
\end{align}
The soft photon, high energy limit of the cross section is zero. There is no ``infrared catastrophe".

\section{Conclusion \label{section:conclusion}}

We have shown that the final state radiation process $e^{+}e^{-}\rightarrow\mu^{+}\mu^{-}\gamma$ is not associated with any IR (or UV) divergence at tree level in the soft photon high energy limit if the computations are carried out with a careful treatment of a distributional object associated with the Feynman amplitude for the process. The soft photon high energy limit of the cross section for the final state photons for this process is found to be zero. There is no ``infrared catastrophe". There are two main reasons for this IR finiteness. The first is the cancellation that occurs between lines 6 and 7 of the sequence of equations given by Eq. \ref{eq:miraculous_cancellation}. The second is the aforementioned distributional representation of an object associated with $\overline{|{\mathcal M}|^2}$ in the high energy limit.

With more work, the computation could be done at arbitrary energies of the leptons and photons involved in the process but the computations become very complicated. If this were done then one could compute the distribution of final state photon energy from the differential cross section for the total process.

\end{document}